\documentclass[conference]{IEEEtran}
\IEEEoverridecommandlockouts
\usepackage{cite}
\usepackage{amsmath,amssymb,amsfonts}
\usepackage{algorithmic}
\usepackage{graphicx}
\usepackage{textcomp}
\usepackage{xcolor}
\def\BibTeX{{\rm B\kern-.05em{\sc i\kern-.025em b}\kern-.08em
    T\kern-.1667em\lower.7ex\hbox{E}\kern-.125emX}}
\usepackage{url}
\usepackage{subcaption}
\usepackage{todonotes}

\begin{document}

\title{A Vision for Flexibile GLSP-based Web Modeling Tools}

\author{\IEEEauthorblockN{Dominik Bork}
\IEEEauthorblockA{\textit{Business Informatics Group, TU Wien} \\
Vienna, Austria \\
dominik.bork@tuwien.ac.at}

\and

\IEEEauthorblockN{Philip Langer}
\IEEEauthorblockA{\textit{Eclipsesource}\\
Vienna, Austria \\
planger@eclipsesource.com}

\and

\IEEEauthorblockN{Tobias Ortmayr}
\IEEEauthorblockA{\textit{Eclipsesource}\\
Vienna, Austria \\
tortmayr@eclipsesource.com}
}

\maketitle

\begin{abstract}
In the past decade, the modeling community has produced many feature-rich modeling editors and tool prototypes not only for modeling standards but particularly also for many domain-specific languages. More recently, however, web-based modeling tools have started to become increasingly popular for visualizing and editing models adhering to such languages in the industry.
This new generation of modeling tools is built with web technologies and offers much more flexibility when it comes to their user experience, accessibility, reuse, and deployment options. One of the technologies behind this new generation of tools is the Graphical Language Server Platform (GLSP), an open-source client-server framework hosted under the Eclipse foundation, which allows tool providers to build modern diagram editors for modeling tools that run in the browser or can be easily integrated into IDEs such as Eclipse, VS Code, or Eclipse Theia.
In this paper, we describe our vision of more flexible modeling tools which is based on our experiences from developing several GLSP-based modeling tools.
With that, we aim at sparking a new line of research and innovation in the modeling community for modeling tool development practices and to explore opportunities, advantages, or limitations of web-based modeling tools, as well as bridge the gap between scientific tool prototypes and industrial tools being used in practice.

\end{abstract}


\section{Introduction}
\label{sec:intro}
Efficient techniques and platforms for the development of modeling languages and tools, such as language workbenches and meta-modeling frameworks, have been a research endeavor in the modeling community since decades~\cite{Kelly.96,steinberg2008emf,Jarke.95}. This is not surprising, because many of the innovative contributions of this community, such as new domain-specific languages or algorithms to process or transform models, only 'come to life' with appropriate tool support. This tool support is not only essential for properly evaluating the feasibility and characteristics of the proposed approaches, but also for sparking new research initiatives around this topic, allowing others to build upon existing work. This is why, for instance, the MODELS and the ER conference series, two of the premier outlets for cutting-edge modeling research, offer dedicated tool tracks to encourage researchers in sharing their modeling tools alongside the theory and conceptual work. Historically, tool development workshops date back to even 2010~\cite{Ossher.10}.

In recent years, there has been a trend towards migrating software development tools (IDEs) to web-based applications and making them available as a cloud service. Prominent examples of this are Github Codespaces and the transition from Visual Studio to VS Code.
This move to the cloud was only recently transferred to the development of modeling tools. While the strengths of web and cloud-based modeling tools are undisputed compared to the traditionally heavy-weight desktop modeling tools~\cite{Rodriguez-Echeverria18a,Rodriguez-Echeverria18}, research into their development, deployment, and operation is still in its infancy. 
With Eclipse GLSP and emf.cloud, the Eclipse foundation has provided first important technologies for the development of cloud-based modeling tools from scratch and made them available as open source libraries. 
With this vision paper, we aim to spark a new line of research that explores and utilizes the many flexibility options enabled by this new breed of GLSP-based web modeling tools.

The remainder of this paper is organized as follows. Section~\ref{sec:background} provides background information on how GLSP uses the concepts of LSP to support editing of graphical diagrams (i.e., models). The need for flexible web modeling tools is discussed in Section~\ref{sec:need4flexibility}. Section~\ref{sec:glsp-flexibility} then elaborates on the many flexibility options offered by GLSP-based web modeling tools. 
Our vision toward flexible modeling tools is summarized in Section~\ref{sec:vision} before we conclude this vision paper in Section~\ref{sec:conclusion}.

\section{Background}
\label{sec:background}
In the following, we will briefly establish the relevant foundations necessary to understand the workings of GLSP. 

\subsection{Language Server Protocol}
\label{sec:background:lsp}
In industry, tool providers are striving for making tool development increasingly efficient. A modern and popular example of such an endeavour is the Language Server Protocol (LSP), which evolved to be the de-facto standard for developing language editing support in modern Integrated Development Environments (IDEs), since it was originally introduced by Microsoft, RedHat, and Codeenvy in 2016~\cite{bunder2019decoupling}.

The core idea behind LSP is to split the traditionally rather monolithic language implementations of IDEs into a language client, which is a user-facing generic editor, and a language server, which encapsulates the implementation of the language smarts of a language, such as parsing, indexing, and refactoring support, in an IDE-independent backend component. The protocol of LSP itself standardizes the communication between these two components so that the client only needs to be able to interpret and understand the protocol instead of the specific programming language, whereas the server can focus on the language support but does not need to consider the specifics of the respective IDE. This reduces the complexity of realizing language support on different IDEs and opens up the development of language support in an arbitrary programming language, independently of any IDE into which the language shall be integrated.
Currently, version 3.17 of the protocol describes 40 different messages between client and server and has an implementation for over 100 different programming languages/technologies~\cite{microsoftlspimpl,microsoftlspspec}.

\subsection{Graphical Language Server Protocol}
\label{sec:background:glsp}
\begin{figure*}[t]
    \centering
    \includegraphics[width=.95\linewidth]{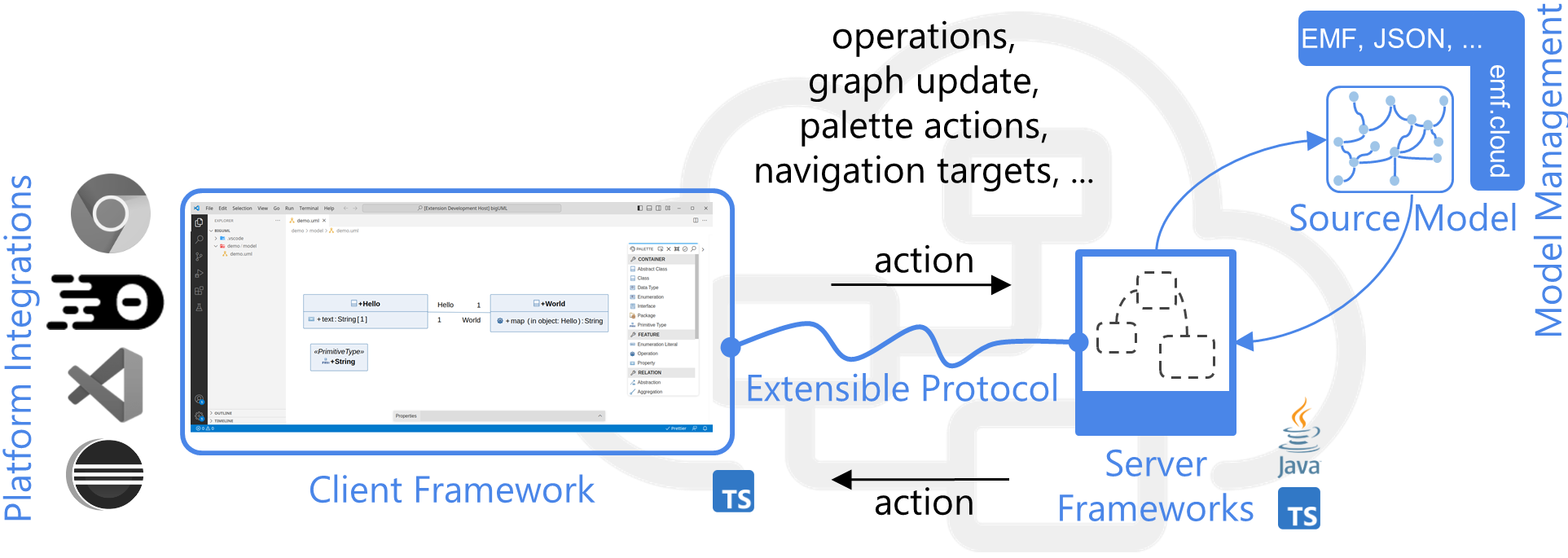}
    \caption{Overview of GLSP components and their interplay.}
    \label{fig:glsp-overview}
\end{figure*}

Initially, LSP has only been defined and used for text-based languages. Still, it was quickly discovered that this concept could also be applied to other areas, one of them being graphical languages by the research community~\cite{Rodriguez-Echeverria18} and the open-source modeling community at Eclipse~\cite{firstglsp}.
Soon after, the Eclipse Graphical Language Server Platform (Eclipse GLSP)~\cite{eclipseglsp} has been established as an open-source project that uses an LSP-like protocol, as well as generic framework components to enable the development of custom, web-based diagram editors, transferring LSP's client/server architecture for diagrams (see Figure~\ref{fig:glsp-overview}).
Thus, the server is responsible for model management, editing logic, validation, and manipulation of the underlying model(s) and communicates via a JSON-RPC (web-)socket to a client, which is responsible for rendering the graphical representation of a model and handling user interactions.
Besides a defined set of protocol message types, the communication between the client and the server is centered around a \emph{graphical model}, which is shared between the client and the server and which describes the hierarchical structure and state of a diagram based on an attributed, typed graph on a two-dimensional coordinate system.
On the client, this graphical model is rendered as an SVG element inside a browser with the help of Eclipse Sprotty\footnote{\url{https://github.com/eclipse/sprotty} (Accessed: 16.05.2022)}. 
User interactions on this SVG graph on the client may result into GLSP actions, which are, depending on their type, either handled locally on the client, e.g., for panning, zooming, or visual feedback, or they are transferred back to the server, e.g., to perform a manipulation of the underlying model(s).
If an action on the server results in a model change that affects the diagram, the server processes the change in its internal model management and eventually sends a new version of the graphical model to be rendered back to the client to refresh the diagram view.

GLSP is under active development by the community since 2017, with the next major release v2.0.0 (expected July 2023). In its current version, GLSP provides four types of components for realizing modern web-based modeling tools (see Figure~\ref{fig:glsp-overview}): 
\begin{itemize}
    \item \textbf{Server framework.} GLSP provides a server framework one can use to build particular diagram servers for e.g., UML or a domain-specific graphical modeling language on top of. Initially, GLSP was focused on supporting the Eclipse Modeling Framework (EMF), based on which many modeling languages and their language-specific logic are already implemented and GLSP servers have been mainly written in Java. In the meantime, this support opened up to arbitrary model management frameworks, whether it is EMF, a JSON file, a database or a remote REST service. More recently, GLSP also added a framework that enables the development GLSP servers with TypeScript.

    \item \textbf{Client framework.} GLSP also provides a client framework. Similarly to the server framework, one can build a particular graphical modeling language client including the definition of the rendering with SVG, styling, and user interaction on top of the provided GLSP client framework. As the rendering and user interaction may heavily differ between one graphical modeling language and another, the client framework allows users to take full control over the SVG view implementations for rendering and enable the customization, as well as adding additional editing tools to control user interaction.
    
    \item \textbf{Protocol.} The messages that can be exchanged between the GLSP clients and servers are specified in a flexible and extensible GLSP protocol which standardizes, in a language-agnostic abstraction level, the communication between arbitrary clients and servers.
    
    \item \textbf{Platform integration.} GLSP provides platform integrations, reusable components that take an implemented GLSP diagram client and integrate it seamlessly into platforms such as Eclipse RCP, Eclipse Theia, or VS Code. These components provide the clue code necessary to register an editor to a certain file type or some other commands specific to the integrated platform. With that, GLSP aims at enabling the integration of GLSP editors into multiple tool platforms and application with maximum reuse.
\end{itemize}


\section{On the Need for Flexibility}
\label{sec:need4flexibility}

The idea behind LSP and GLSP is not just driven by the goal of migrating (modeling) tools towards a web-based UI technology stack.
It aims also at breaking with monolithic architectures, tight coupling with underlying tool platforms and fixed deployment architectures (such as a desktop client or a cloud-based deployment).
In fact, the tool market in the recent years has significantly changed.
Instead of a single heavy-weight monolithic desktop applications that needs to support all possible tasks along the software development process, users expect their language-specific tools to be compatible and only lightly integrated with multiple tool platforms or editors such as VS Code, Eclipse Theia, VI, or Emacs.
This enables a more flexible combination of the best-in-class components.
Also more development tools need to support both, running as a desktop application operating on a local filesystem, but also being deployed in the cloud based on ephemeral, task- or context-specific workspaces.
Users expect to click a link, for instance on a pull request, and end up in a prepared and readily configured tool, running in the browser, where the changes of the pull request are already checked out, possible pre-built, and all runtime components, such as those required for debugging or generating and running code, are available in the cloud container, while still being able to edit the project at hand with full editing support of a powerful IDE running in the browser.
Thus the components working on those diverse environments must be agnostic and loosely coupled with the filesystem and their runtime (browser, cloud-infrastructure, or desktop).
Any change applied by the user in this environment is eventually put into a new commit on that pull request to store the state back into the repository.
After a task is completed, the workspace is thrown away and a new one is created for the next task at hand.
This trend, which increasingly becomes the state of the art in software engineering, slowly arrives also in the world of modeling tools, too. 

Flexibility with respect to the modeling approach itself, is not a new topic, see~\cite{Rose.12,Guerra.18}. There even have been scientific workshops in the past that were dedicated to flexible tools~\cite{Ossher.10}. 
Within the modeling community, flexibility is often related to informal vs. formal modeling~\cite{Bork.20,Gabrysiak.11,Harel.2000syntax,Zarwin.2012} or the availability of multiple, stakeholder-specific concrete syntaxes (cf.~blended modeling~\cite{Glaser.21,David.23}). 
Many works focus on mitigating the dichotomy~\cite{Ossher.10,Rose.12} between very formal and powerful tools supporting experts in later stages of engineering projects with the informal, flexibly, and creativity-fostering tools that can be used by non-experts in the earlier stages~\cite{Sandkuhl.18,Wuest.19}. 

There is only very few research that concerns the flexibility of language workbenches or metamodeling platforms. In most of the papers reporting of such platforms, the term 'flexible'~\cite{Atkinson.16,Smolander.91} refers to the flexibility of $i)$ supporting arbitrary modeling languages, and $ii)$ realizing domain-specific concrete syntaxes. 
FlexiSketch~\cite{Wuest.19} is an example of tools that allow sketching models, i.e, using an informal approach to create early designs which are then later translated into formal models adhering to a formal modeling language.

What is lacking so far, and what we propose in this vision paper, is to move the flexibility discussion to the platforms we use to develop modeling tools, and consequently, to the flexibility entailed in the deployed and used tools themselves. 
In the next section, we will basically take this idea of flexibility and walk through the GLSP architecture and its components (cf.~Section~\ref{sec:background:glsp}), and give some examples of where flexibility is crucial and how it is enabled by GLSP.

\section{Flexibility of GLSP-based Web Modeling Tools}
\label{sec:glsp-flexibility}

Over the years, the community around GLSP pushed towards making several aspects of GLSP more flexible and versatile across all components of GLSP. 
In the remainder of this section, we summarize and reflect on the use cases and flexibility that has been added to GLSP by the community based on industry needs.

\subsection{Flexibility via Inversion of Control}

One of the main goals of the GLSP initiative is to provide a flexible and reusable framework for building graphical modeling tools that can be easily integrated into any application frame and deployment scenario. Every aspect of the framework might need to be extended or customized. Therefore, following the example of modern web-based tools such as Eclipse Theia and Sprotty, an \textit{Inversion of Control} pattern based on dependency injection is used. Each GLSP component encapsulates implementation logic and services into reusable feature modules. Modules can be extended or customized toward the needs of a specific use case. This facilitates a composable architecture of reusable and interchangeable components. 
We observe that the flexible, modular approach has become one of one the main arguments for adopting GLSP by the industry because adopters can fully tailor the framework to their needs.

\subsection{Flexibility on the Client}
\label{sec:flexibility:client}
Recently, language workbenches and frameworks for building modeling tools, such as Eclipse GEF, GMF runtime, GMF tooling, Sirius, were building an increasing number of abstraction layers hiding the details of the actual user interface implementation. While this arguably makes it very simple to get started, as technical details are hidden away, this also hampers the power of tool developers to carefully design the user experience, often leading to generic and poor usability, as was also reported by adopters of that technologies~\cite {8101269}.
With the modern, web-based user interface technology stack, including HTML5, CSS, and SVG, new opportunities for rethinking the dusty usability concepts of traditional modeling tools arise.

\begin{figure}[htb!]
\vspace{-.2cm}
\begin{subfigure}[t]{0.24\columnwidth}
\centering
\includegraphics[width=\textwidth]{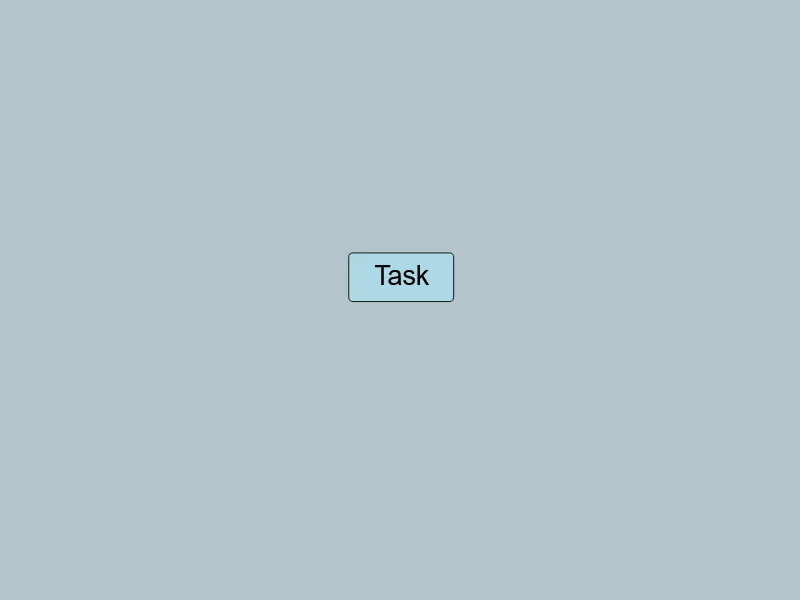}
\subcaption{}
\label{fig:prototype1:a}
\end{subfigure}\hfill
\begin{subfigure}[t]{0.24\columnwidth}
\centering
\includegraphics[width=\textwidth]{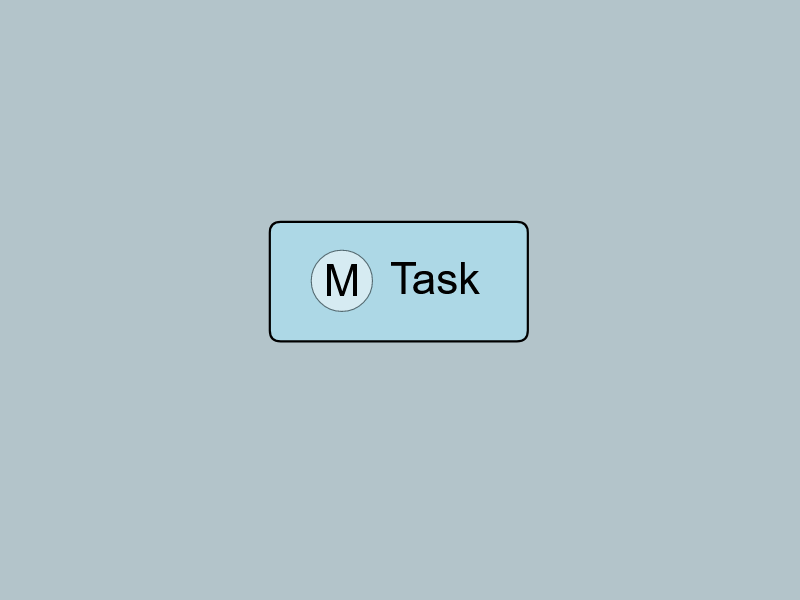}
\subcaption{}
\label{fig:prototype1:b}
\end{subfigure}\hfill
\begin{subfigure}[t]{0.24\columnwidth}
\centering
\includegraphics[width=\textwidth]{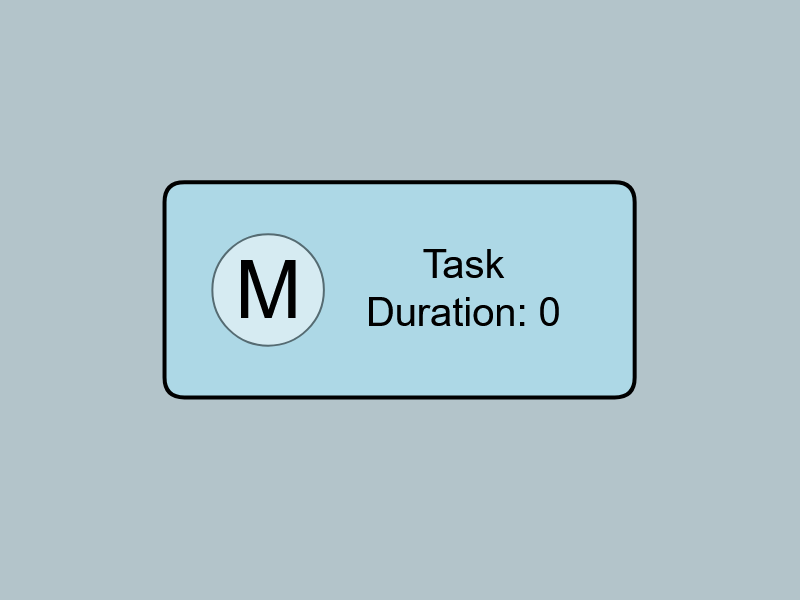}
\subcaption{}
\label{fig:prototype1:c}
\end{subfigure}\hfill
\begin{subfigure}[t]{0.24\columnwidth}
\centering
\includegraphics[width=\textwidth]{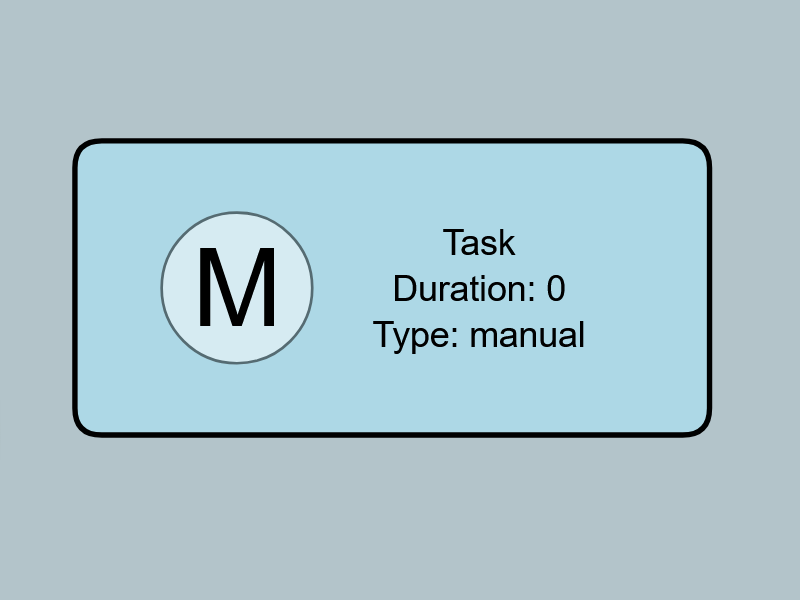}
\subcaption{}
\label{fig:prototype1:d}
\end{subfigure}
\vspace{-.2cm}
\caption[]{Dynamic GLSP-client side rendering\footnotemark.}
\label{fig:prototype1} 
\vspace{-.3cm}
\end{figure}

With the increasing adoption of Eclipse GLSP in industry, we increasingly observed the force driven by industrial use cases to avoid burying the user interface implementation below layers of abstraction, as was often the case in traditional modeling frameworks, but rather empower tool developers with direct access to the excellent and well-known UI technologies to give them full control over look and feel of their modeling tools.
This certainly does not mean having to build everything from scratch.
Instead, the reusable user interface concepts are provided as a library of shapes, editing tools, UI components, etc., which can be used as is, but also be customized or even replaced entirely.
Consequently, in GLSP, the SVG generation and CSS styles for diagrams are not hidden anymore behind abstract diagram configurations, but are directly exposed in the form of reusable library components, which are open to modification in order to give full flexibility in realizing advanced model representation and the user interaction by means of editing tools (cf.~\cite{DeCarlo.22-er} for a taxonomy of advanced features). This decision also enables to benefit from the rich set of available experience of working with languages like SVG but also to use the availability of excellent debugging tools for these languages.
Besides the SVG view implementations, adopters of GLSP can even fully customize the editing tools of the modeling tool to account for the particular needs of a domain-specific user group and introduce features, such as highlighting the valid target elements after the user selected the source element of a newly created edge, etc.
The generation of SVG can even be highly dynamic (a gallery of examples is provided online\footnote{\url{https://www.eclipse.org/glsp/gallery/}}). Recent research showed, how the standardized mapping of a graph model element to an SVG element can be flexibly extended by dynamically adjusting e.g., the rendered $i)$ form, and $ii)$ content of the elements based on the currently visible \textit{zoom level}~\cite{DeCarlo.22-models}, see Fig.~\ref{fig:prototype1} for an illustration of the dynamic adaptation of the rendered model in the GLSP client contingent on the current zoom level.

\footnotetext{Semantic Zoom video: \url{https://www.youtube.com/watch?v=iBs-fGwq15Y}}

Lastly, several use cases raised the need for adding additional UI controls, implemented in plain HTML, CSS and JavaScript, on top of the diagram.
This allows to add controls, e.g., for enabling or disabling certain filters or for editing certain aspects of the model with web forms or complex text boxes after, for instance, selecting a diagram element.

\subsection{Flexibility on the Integration}
\label{sec:flexibility:integration}
Based on our experience, flexibility also matters when thinking of integrating the GLSP-based diagram editors into a tool platform like Eclipse, VS Code, or Eclipse Theia. With GLSP, there is a clear separation of concerns in place where the diagram editor is kept entirely platform-independent and only integrated via platform-specific glue code, which in turn interfaces to the platform's native APIs. From this separation of concerns follows that a GLSP tool developer can not only use the diagram editors with maximum reuse across multiple tool platforms but also benefit from the full power of the platform's native APIs to implement a seamless integration between the editor and the tool platform, including populating error markers to the platform's problems view, allowing navigating across views and editors to diagram elements, etc.

\begin{figure}[ht!]
    \centering
    \includegraphics[width=.9\linewidth]{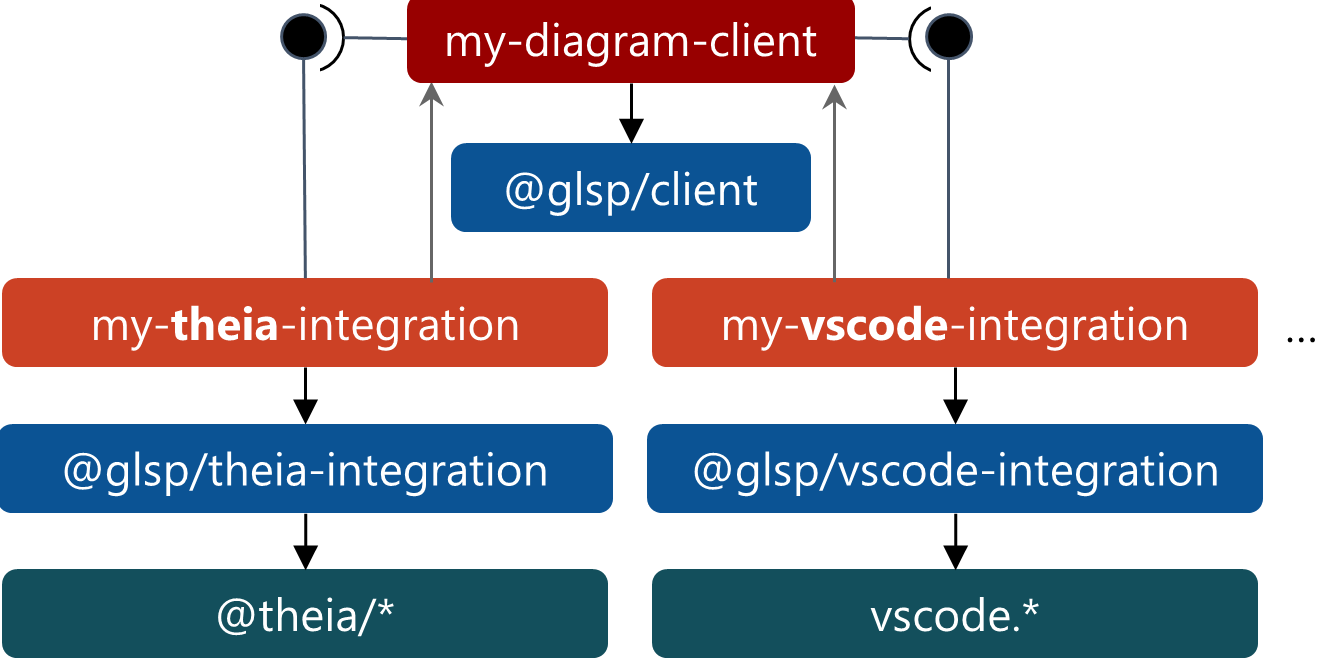}
    \caption{Flexible platform integration of GLSP diagram editors.}
    \label{fig:integration}
\end{figure}

The architecture of this separation is schematically illustrated in Fig.~\ref{fig:integration} for Theia and VS Code. The GLSP diagram client exposes an interface to which an extended and customized Theia or VS Code integration, which builds on the native Theia or VS Code integration, needs to bind. 
These integrations allow replication for other platforms. The flexibility here is just as important as the diagram itself because a good workflow and a good modeling tool do not start and end at the borders of a diagram editor, but rather span the entire flow when the user opens the tool and works through their process throughout the tool.

\subsection{Flexibility in the Protocol}
\label{sec:flexibility:protocol}
A core enabler of the flexibility of GLSP-based modeling tools is the protocol that is being used to coordinate the GLSP clients with the GLSP server. This protocol shares the fundamental idea behind the Language Server Protocol (LSP) and is designed to cover the most common actions adhering to graphical diagram editors out of the box.
As diagram editors are usually very specific in their interaction and capabilities though, it is worth noting that the protocol is intended to be enhanced with custom actions, which can either be sent from the client to the server to, for instance, execute some model processing on the source model or to notify the server about UI events in custom UI controls.
Likewise, the server can define custom actions to send to the client, e.g., to inform the client about specific validity checks or about additional domain-specific information.

Another flexibility enabled by the protocol is that it abstracts away from the underlying technologies used to implement the GLSP client and GLSP server. This enables the development of further client and server frameworks in different technologies -- something we see now with the new TypeScript-based GLSP server framework alongside the existing one for Java. 
Here is where we see that GLSP further inherits the strengths of the LSP where we recognize an increase of available LSP clients\footnote{\url{https://microsoft.github.io/language-server-protocol/implementors/tools/}}, LSP servers\footnote{\url{https://microsoft.github.io/language-server-protocol/implementors/servers/}} and LSP Software Development Kits\footnote{\url{https://microsoft.github.io/language-server-protocol/implementors/sdks/}}. 
It would be surprising to not see similar developments surrounding GLSP in the near future. 

\subsection{Flexibility in the Model Management}
\label{sec:flexibility:modelmanagement}
Flexibility in model management was a huge objective with the 1.0 release of GLSP. It is now the developer's choice how to realize model management, i.e., to decide \textit{which format} to use, which \textit{framework} to use, whether it is \textit{local or remote}, and whether it is realized as a \textit{shared service} across users or \textit{isolated}. 
This flexibility certainly entails additional implementation effort for specific model management frameworks, as model commands, loading and saving models is inherently specific to the used model management. To mitigate this additional effort, GLSP with its 1.0 release also pulled out the generic implementations into resuable model management modules to simplify writing GLSP servers that interact with EMF, Json, and emf.cloud\footnote{\url{https://github.com/eclipse-emfcloud/emfcloud-modelserver}}. 
This flexibility not only supports the migration of existing tools into GLSP, but also prevents a lock-in in the future. 

Eventually, this also enables model management reuse across multiple deployments and in different platform integrations. 
For example, a local tool can interact with the local file system while when integrated into a web application, the same tool, with just a replaced model management module, can use a shared model management or a shared model server where several people can have read/write access. 

\begin{figure}[ht!]
\vspace{-.3cm}
   \centering
   \includegraphics[width=.99\linewidth]{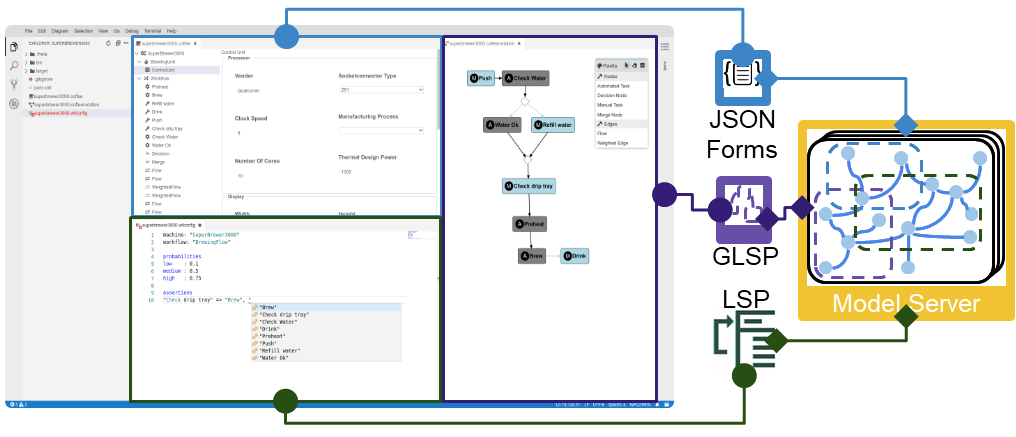}
   \caption{Flexibility in model management.}
   \label{fig:modelmanagement}
\vspace{-.3cm}
\end{figure}

GLSP comes now with the emf.cloud model server component\textsuperscript{7} which enables multiple widgets in your tool interact with the same underlying model. The emf.cloud model server essentially provides a component that loads and manages these runtime states of the model so that they can be then interacted with from different widgets with different services like GLSP editors, Json forms, and LSP editors (see Fig.~\ref{fig:modelmanagement}). 

\subsection{Flexibility on the Server}
\label{sec:flexibility:server}
Due to the clear protocol-based separation between the client and the server, tool developers can choose to write their GLSP servers in any programming language as long as they adhere to the defined protocol.
However, writing servers from scratch entails quite some effort.
GLSP, therefore, provides server frameworks that already cover all generic features and also provides supporting libraries to implement the diagram-specific functionality.

At its inception, GLSP only provided a Java-based server framework. With GLSP 1.0, however, a framework for TypeScript has been added, which gives tool developers the flexibility to choose what is the best fit for their project without the penalty of having to implement a server from scratch.
The TypeScript-based server framework has significant advantages for use cases where one is targeting a VS Code or non-cloud-based Theia platform, as these already come with the nodejs runtime and, thus, don't entail any additional runtime requirements, such as a Java Virtual Machine, on the user's machine.
Moreover, a TypeScript-based GLSP server leads to a more homogenous development stack alongside the TypeScript-based client.

\subsection{Flexibility in Deployment}
\label{sec:felxibility:deployment}
\begin{figure}[ht!]
    \centering
    \includegraphics[width=\linewidth]{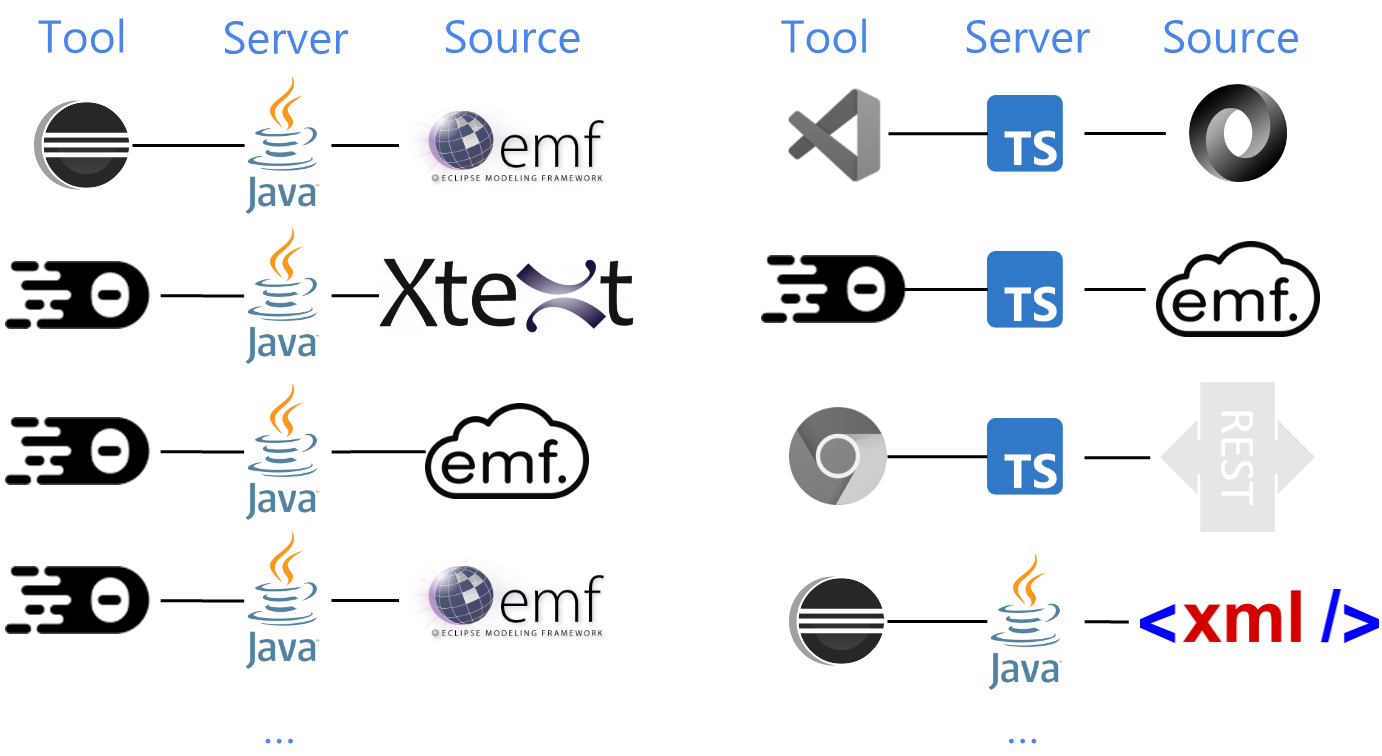}
    \caption{Flexibility in deploying GLSP-based modeling tools.}
    \label{fig:glsp-deployment}
\end{figure}

The GLSP architecture and the flexibilities discussed at the outset, especially on the server and the model management, enable high flexibility with respect to the deployment of GLSP-based modeling tools. Fig.~\ref{fig:glsp-deployment} illustrates common deployment options observed in the industry. Note that this illustration is not aimed to be comprehensive and further options are likely possible. The deployment options are clustered along three decisions: $i)$ which tool integration to use?; $ii)$ which GLSP server framework to use?; and $iii)$ which model management framework to use. 

From a runtime perspective, one can further decide whether the components run in their own, separate processes, in one single process, or even in separate containers, which is particularly valuable in a cloud infrastructure scenario. 
While in Eclipse it makes sense to have everything running in one process if your server is Java-based, in VS Code extensions typically run a separate process running in nodejs. In a Theia cloud deployment, one can deploy a separate Docker container and even go as far as extracting the model management into its own potentially shared container.
Eventually, having a TypeScript-based server even enables lightweight deployment scenarios, in which no separate process and no cloud infrastructure is required at all and the entire GLSP editor, including the client and the server, is running in the browser only.

\section{The Vision for Flexible Web Modeling Tools}
\label{sec:vision}
This vision paper aims to spark attention to the possibilities offered by highly-flexible platforms like Eclipse's Graphical Language Server Platform (GLSP) and the web modeling tools build with them. 
The modeling community has a long tradition of developing modeling tools. However, often these innovative tools remain on a scientific prototypical level~\cite{Pourali.18} and only a few of them, e.g., Papyrus~\cite{Lanusse.09papyrus}, reach a maturity level that qualifies for wide-spread educational and industrial application.

We believe with the uptake of platforms like GLSP it is now time to move forward as a community and bridge the gap between academic prototypes and modern industrial tools. With the flexibility of this new breed of modeling tool development platforms, researchers and engineers are now able to start from standardized base (web) technologies to realize state-of-the-art tools, but still have full control over many aspects, from usability to model management, to explore and facilitate innovative approaches to modeling languages and tools.
The use of standard open-source technologies should also account for sustainability and mitigate the lock-in effect which is particularly harmful in a scientific environment where PhD students and Postdocs, who often drive the development of a tool prototype, leave academia. 

An open and interesting endeavor toward the realization of this vision is to bring the existing tools and the vast amount of powerful EMF-based technologies to this web-based modeling technology stack.
For GLSP, this problem is mitigated in so far, as with the EMF model server in emf.cloud, there is a component ready to use that can be likely equipped and extended with the power of the EMF ecosystem.
Still, there is lots of room for research and innovation towards easing the transition, providing modern and responsive front-ends to existing EMF technologies, such as EMF Compare, model query and transformation approaches, model refactoring, and many great achievements of the modeling community.
For all those approaches, we would be thrilled to see not only a one-to-one migration but also a re-evaluation of how the usability and flexibility of those approaches can be enhanced based on the power of this new web-based technology stack.

To learn from existing solutions and to not re-do the same mistakes from the past, it would be also essential to establish a broader open knowledge base of best practices and successful tool developments with GLSP. With the further maturing of the technologies and the increasing attention by the community, we expect this is just a matter of time. 
Given that GLSP has this clear separation of concerns, such a knowledge base or a source code repository would greatly foster reuse across modeling tools. 
Therefore, existing high-quality solutions can be easily injected and reused for common requirements while unique features of a specific modeling tool can still be customized to the highest extent. 

The open and flexible architecture of these new breeds of modeling tools also enables the efficient injection of external functionality and features that are also built on base web technologies. For example, existing frameworks for testing the accessibility of web applications can now be adapted to test modeling tools with respect to their accessibility for modelers with disabilities -- a topic mostly ignored until now with increasing importance. 
Moreover, the open technology stack also eases the reuse and injection of AI/ML solutions to support downstream conceptual modeling tasks like model pattern discovery~\cite{fumagalli2022pattern}, domain classification~\cite{rubei2021lightweight,lopez2022machine}, model completion~\cite{burgueno2021nlp}, refactoring~\cite{lahijany2021identibug}, repair~\cite{fumagalli2020towards}, and model transformation~\cite{burgueno2022generic}. All this should ease the realization of smart modeling assistants~\cite{Mussbacher.20}.

To realize this vision, it is essential, that the modeling community further strengthens the links to industry and open-source communities, such as the Eclipse GLSP community. 
By increasing collaborations, researchers can gain access to the needs of the industry while the platform vendors can learn from the innovative approaches developed in academia and support translating them into common platform features. 

\section{Conclusion}
\label{sec:conclusion}
In this paper, we envisioned the future of much more flexible web modeling tools. Using the Graphical Language Server Protocol (GLSP) platform as the currently most promising modeling tool development platform, we elaborated in detail on the many facets of flexibility that are enabled by this new breed of tool development platforms. 
Obviously, our presentation is focused on the concrete flexibility enabled by GLSP. However, as GLSP heavily relates to standard web technologies and communication protocols and architecturally aligns with state-of-the-art approaches for textual language engineering (LSP), we believe many of the presented flexibilities can be translated to other, future tool development platforms we are not aware of today. 
By moving from proprietary technologies toward standardized open-source technologies, the tools developed by the MODELS community can eventually bridge the current gap between academic prototypes and modeling tools used in industry. 
This may also foster collaborations between the modeling research community and practice (e.g., tool vendors).

\bibliographystyle{IEEEtran}
\bibliography{IEEEabrv,mybibfile}

\vspace{12pt}

\end{document}